# An on-chip programmable mechano-quantum transducer

Xinrui Zhang[1,#], Wei Liu[2,5,#], Duanyu Ma[1], Lin-Ke Xie[2,5], Nai-Jie Guo[2,5], Zhongtao Gou[1], Yifan Wang[1], Jianxin Xu[1], Xiaoguang Luo[3], Zhao Mu[1], Honglong Chang[1], Weizheng Yuan[1], Jian-Shun Tang[2,5,6]*, Chuan-Feng Li[2,5,6]*, Guang-Can Guo[2,5,6], Tao Ye[1,4]*

[1]Ministry of Education Key Laboratory of Micro/Nano Systems for Aerospace, Key Laboratory of Micro- and Nano-Electro-Mechanical Systems of Shaanxi Province, School of Mechanical Engineering, Northwestern Polytechnical University; Xi'an, 710000, China

[2]Key Laboratory of Quantum Information, University of Science and Technology of China, Hefei, 230026, China

[3]State Key Laboratory of Flexible Electronics (LOFB) & Institute of Flexible Electronics (IFE), Northwestern Polytechnical University, Xi'an, 710000, China

[4]Key Laboratory of Scale Manufacturing Technologies for High-Performance MEMS Chips of Zhejiang Province, Key Laboratory of Optical Microsystems and Application Technologies of Ningbo City, Ningbo Institute of Northwestern Polytechnical University, 218 Qingyi Road, Ningbo, 315103, China

[5]Anhui Province Key Laboratory of Quantum Network, University of Science and Technology of China, Hefei 230026, China

[6]Hefei National Laboratory, University of Science and Technology of China, Hefei 230088, China

[#]These authors contributed equally to this work.

*Correspondence should be addressed to Tao Ye (email: yetao@nwpu.edu.cn), Chuan-Feng Li (email: cfli@ustc.edu.cn), and Jian-Shun Tang (email: tjs@ustc.edu.cn).

## Abstract

Solid-state spin defects encode local perturbations as measurable shifts in spin-transition frequencies, but mechanical actuation and quantum readout remain physically separated, resulting in a discrete measurement setup. Integrating these functions requires an on-site mechano-quantum interface that programs the lattice state of a defect host and quantitatively maps it onto the spin Hamiltonian. Here we first report an on-chip programmable mechano-quantum transducer (OCPMQT) that integrates voltage-defined micromechanical actuation with in situ spin-frequency readout in a two-dimensional van der Waals quantum-defect host. Mechanically programmed lattice states are encoded as shifts in the axial zero-field splitting parameter and resolved by optically detected magnetic resonance (ODMR) spectroscopy. Within a chip volume of $2.05\times10^{-2}$ $cm^3$, the transducer accesses ODMR-inferred strains as low as 0.0080±0.0004% and delivers a volumetric force density of approximately $2.6\times10^4$ $N\cdot m^{-3}$. A micromechanical-to-spin-Hamiltonian framework links on-chip electromechanics, interfacial strain transfer, and strain-spin coupling, enabling the electrical control micromechanical input to be measured directly as spin-frequency response.

## Introduction

The controlled preparation, manipulation, and readout of quantum states underpin advances in quantum information processing[1-3], communication[4-6], and precision measurement[7-9]. Optically addressable[10-15] and coherently controllable[16-22] solid-state spins enable quantum sensing by encoding external perturbations as measurable shifts in spin-transition frequencies. Advancing these systems from spectroscopic probes towards functional devices, however, requires a programmable on-site mechano-quantum interface that maps chip-level mechanical inputs onto quantitative variations in spin Hamiltonian, beyond mere detection of external signals[23,24]. This challenge is particularly pronounced for mechanical perturbations, including pressure[25-28], strain[29-31], and vibration[32-34], which couple to the spin system through deformation of the host lattice.

On-chip micromechanical actuators provide precisely voltage-defined and localized mechanical control, delivering quantifiable forces[35-37] and displacements[38-42] within compact device architecture. Compact micromechanical integration has enabled random-access beam steering on a 1.0×1.1 $cm^2$ silicon-photonic device with sub-megahertz switching[39]. The micromechanical-based actuation platform also supports 4 μm vertical displacement and ±10° rotation of two-dimensional (2D) material interfaces within a 1.0×1.0 $cm^2$ platform[38], as well as twist-free 3.6 μm biaxial positioning that enhances the spatial bandwidth product of image sensors by up to 33.7-fold[40]. These implementations demonstrate micromechanical actuation as a generic device-level mechanical-control layer and reconfigurable motion across diverse chip-scale systems.

Existing quantum sensing studies of mechanical perturbations have largely relied on pre-existing input[25,30], indirect loading[29,31], or bulky external apparatus[26-28], leaving the micromechanical state difficult to on-site control and quantify within a unified and compact device architecture. At the hardware level, nominally compact loading modules still occupy approximately 6 $cm^3$.[43] The atomic force microscopy (AFM)-based systems require footprints of about $3\times10^4$ $cm^3$ or more[44], and commercial piezoelectric nanopositioner typically provide volumetric force densities only on the order of ~$10^3$ $N\cdot m^{-3}$.[45] Among mechanically responsive quantum systems, negatively charged boron-vacancy ( $V_B^-$ ) spins in hexagonal boron nitride (hBN) flakes are particularly suited to device integration owing to the 2D van der Waals (vdW) host, which has been readily reported in optoelectronic platforms[46,47] and micro- and nanodevices[48,49]. The atomically thin host is readily transferable, patternable, and compatible with heterogeneous on-chip integration, supporting intimate coupling to photonic cavities, waveguides, and functional heterostructures, while co-locating optical access, material interfacing, and spin-based sensing within a common device. Leveraging established heterogeneous integration and chip-scale packaging strategies[50,51], on-chip integrating programmable microelectromechanical actuation with 2D spin-defect host could unify these functions within an active mechano-quantum interface that could be generalized to other mechanically responsive quantum systems.

Here we first report an on-chip programmable mechano-quantum transducer (OCPMQT) that directly interfaces voltage-defined micromechanical actuation with in situ spin-frequency readout. Within a device volume of $2.05\times10^{-2}$ $cm^3$, the transducer achieves a volumetric force density of approximately $2.60\times10^4$ $N\cdot m^{-3}$, and generates controlled $V_B^-$ hosting lattice deformation, resolved by optically detected magnetic resonance (ODMR) spectroscopy as strains spanning 0.0080±0.0004% to 0.230±0.013% and a reproducible modulation of the axial zero-field splitting (ZFS) parameter $D$ of approximately -55 MHz. We further establish a multiscale mechano-quantum transduction framework that connects electromechanics, interfacial strain transfer, and strain-spin coupling. The resulting closed-form $D(V)$ relation quantitatively reproduces the experimental $V^2$ response, predicting the spin-frequency shift directly from chip-scale inputs with approximately 1% deviation. By unifying controlled micromechanical loading, quantum spin Hamiltonian readout, and predictive modelling within a single and compact chip architecture, OCPMQT advances mechanically coupled spin defects beyond passive and discrete sensing towards on-chip programmable mechano-quantum transduction.

## Design of the OCPMQT

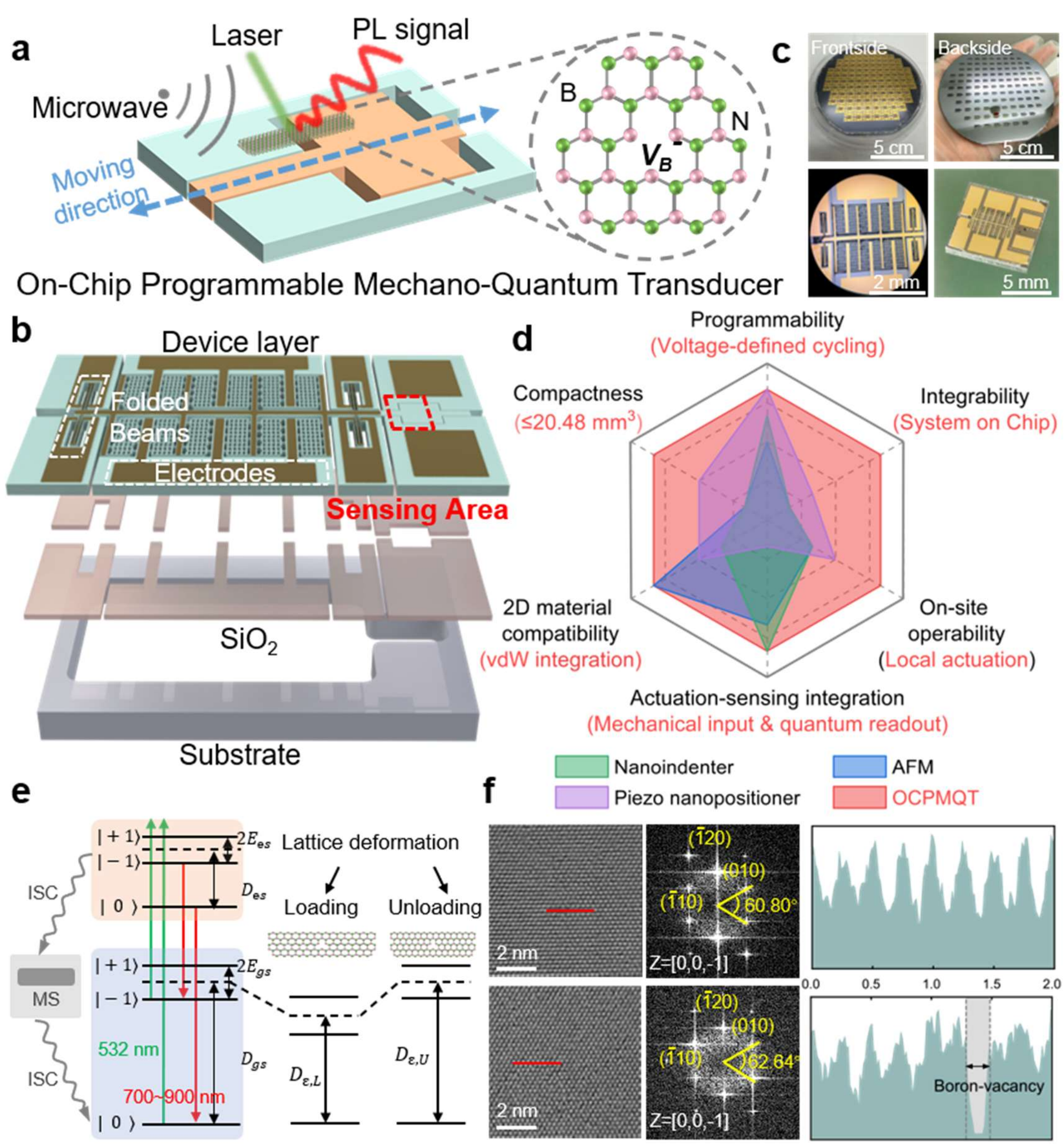

**Fig. 1 | Architecture of OCPMQT for electrically programmable mechano-quantum transduction.**

**a,** Schematic of the OCPMQT. A voltage-driven on-chip micromechanical actuator generates in-plane displacement/force along the moving direction, mechanically loading $V_B^-$ defects in hBN flakes. Optical excitation and microwave control enable photoluminescence (PL) readout from the mechanically coupled defect ensemble. The enlarged schematic illustrates the lattice structure of $V_B^-$ defects in hBN flakes. **b,** Exploded schematic of the main components in OCPMQT. Based on an SOI wafer, comb structures in the device layer convert electrostatic actuation into controlled lateral displacement, enabling micromechanical perturbation in the sensing area. **c,** Photograph of the wafer-scale fabrication of OCPMQT, showing frontside and backside views of the processed SOI wafer, together with enlarged photographs of a diced chip and an individual device. **d**, System-level attributes of the OCPMQT compared with conventional mechanical manipulation platforms, including AFM, nanoindenter and piezoelectric nanopositioner, highlighting compactness, chip-level integrability, voltage-defined cycling, 2D material interfacing, on-site operation and actuation-sensing integration. **e,** Simplified energy levels of $V_B^-$ center and the related radiative transitions (red) and nonradiative ISC transition (blue) among GS, ES, and MS. Tensile loading stretches the lattice and reduces $D_{\varepsilon,L}$, whereas strain release in unloading restores $D_{\varepsilon,U}$ enabling cyclic modulation of the ODMR resonance. **f,** HRTEM characterizations before and after ion implantation, upper and lower diagrams, respectively. FTT diffractograms resolve the crystalline lattice and reveal a change in the lattice-plane angle after implantation. Intensity line scans show atomic contrast loss, consistent with the formation of a boron vacancy.

To realize the control mechanical actuation with in situ spin-frequency readout, we designed an OCPMQT in which programmable micromechanical actuation is coupled to the spin response of $V_B^-$ defects by reversibly loading-unloading the color center lattice and in situ reading out ZFS parameter $D$ shift (Fig. 1a). In this architecture, micromechanical deformation is generated and delivered directly on chip to the quantum material, therefore, establishes a deterministic transduction pathway from chip-scale micromechanical motion to a quantum spin-frequency response. The OCPMQT was fabricated from a silicon on insulator (SOI) wafer with a 30-μm device layer in which in-plane electrostatic actuator, folded beams, and driving electrodes were patterned, followed by vapor HF release of the movable microelectromechanical structures through selective buried oxide layer removal (Fig. 1b,c). The folded beams constrain the movable stage predominantly along the designed $x$ direction, and electrostatic comb drive provides voltage-programmable micromechanical actuation (Supplementary Fig. 1,2). $V_{\mathrm{B}}^{-}$-containing hBN was positioned within the mechanically active region spanning the fixed and movable sections, with oxygen-plasma treatment to enhance adhesion and mechanical coupling to the device layer (Methods). This configuration spatially co-localizes mechanical loading and quantum sensing, thereby directly transferring the microelectromechanical displacement to the defect-hosting lattice. At the system level, the architecture combines voltage-defined mechanical cycling with on-site actuation and sensing in a compact chip-integrable format (Fig. 1d). In contrast to conventional mechanical manipulation techniques, the OCPMQT is designed to integrate mechanical input generation, 2D material interfacing, and quantum readout within a single compact device architecture, establishing a mechanical-writing and quantum-reading strategy for the first time.

The $V_{B}^{-}$ defect in hBN, employed for quantum frequency readout, is one of the most extensively studied spin defects within this material[12]. It is formed by the removal of a boron atom from the hBN lattice, which subsequently captures an electron from the local environment (see inset of Fig. 1a). The corresponding energy-level structure is illustrated in Fig. 1e, where both the ground states (GS) and excited states (ES) are spin triplets ($S$ = 1), accompanied by metastable states (MS) singlet states ($S$ = 0). Within the GS spin triplet, the $|m_s$= 0> and $|m_s$=±1> manifolds are separated by the axial ZFS parameter $D_{\mathrm{gs}}$, while the two branches are further resolved by the transverse ZFS term $2\,E_{\mathrm{gs}}$ [13,47]. Under 532-nm optical excitation, fluorescence emission is detectable in the 700-900 nm range, and the spin population can be polarized into the $|m_s$= 0> state via intersystem crossing (ISC); concurrently, microwave irradiation drives spin transitions from $|m_s$= 0> to $|m_s$=±1>, thereby generating a contrast signal in ODMR. From the two resonance frequencies, $\nu_{-}$and $\nu_{+}$, observed in the ODMR spectrum, the center frequency $D$=($\nu_{-}$+$\nu_{+}$)/2 and the transverse splitting parameter $E$=($\nu_{+}$-$\nu_{-}$)/2 can be derived. For our sample, the ZFS parameters at zero applied voltage are $D_0$ ≈3.51 GHz and $E_0$ ≈71.7 MHz, respectively, with the slightly elevated $D_0$ likely arising from transfer-induced residual strain and the local defect environment.

Consequently, a mechanically induced axial perturbation of the spin Hamiltonian is encoded as a shift of $D$. The defect-hosting lattice was further examined by high-resolution transmission electron microscopy (HRTEM) before and after ion implantation. Following implantation, the thin hBN edge region exhibits reduced atomic-row regularity (Fig. 1f and Supplementary Fig. 3,4). Fast frequency transform (FTT) analysis of selected lattice regions shows a change in the angular relationship between crystallographic planes from approximately 60.80° to 62.64°, and the corresponding intensity profile reveals a local disruption of the periodic atomic contrast, which the observed angular deviation and disruption of the periodic atomic contrast are consistent with implantation-induced lattice perturbation. Figure 1 establishes the integrated programmable actuation architecture and the axial $D$-channel readout. We next examine whether the programmed actuator displacement is mechanically transferred to the defect-hosting hBN.

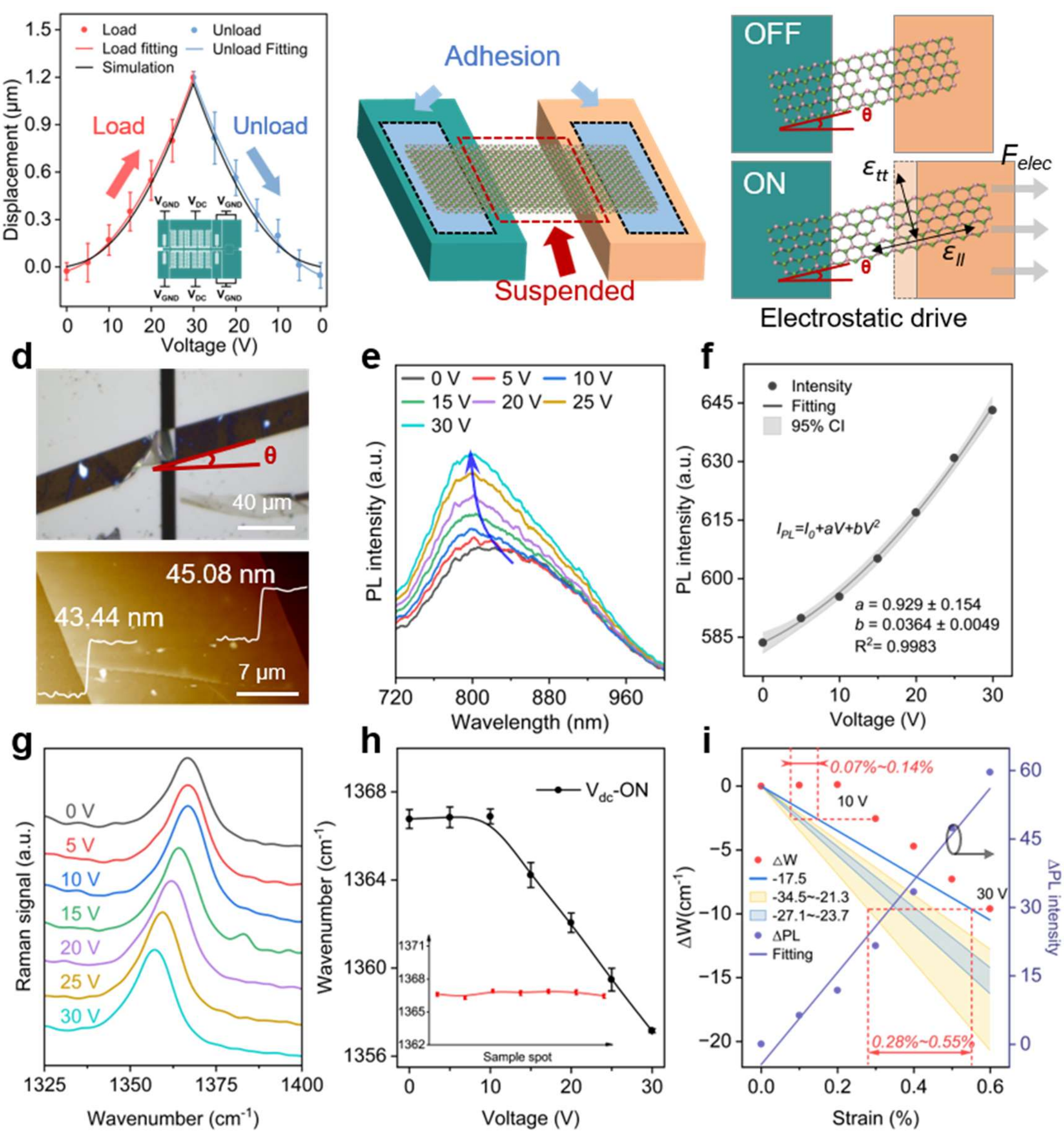

**Fig. 2 | Mechanical loading and optical verification of strain in the OCPMQT.**

**a**, Voltage-dependent displacement of the movable chip region during loading and unloading. The measured reversible displacement is consistent with the simulated distribution. Inset: electrical connection of micromechanical actuation. **b,** Schematic of the hBN flake, which is transferred to the fixed and movable regions, enabling mechanical loading through lateral actuation. **c,** Schematic of sensing area. The hBN flake is placed at an angle $\theta \approx 17°$ with OFF and ON states. Under electrical actuation, lateral displacement stretches the lattice, inducing in-plane tensile strain. **d**, The optical image of hBN and thickness measured by AFM. **e**, PL response of the $V_B^-$ to electromechanical loading, showing a blue-shift of the characteristic peak as the voltage increases. **f**, PL intensity as a function of voltage, fitted by a voltage-dependent response curve. **g-h**, Raman spectra measured at different voltages, showing a red-shift of the characteristic peak as the voltage increases. **i**, Correlation between Raman peak shift, PL-intensity variation, and estimated tensile strain. Reported Raman-strain coefficients define the arrowed strain ranges from minimum strain-10 V and maximum strain-30 V, over which the PL-intensity variation increases approximately linearly.

A mechano-quantum transducer requires the programmed microelectromechanical motion to reach the defect-hosting lattice rather than remain solely as free actuator displacement. We therefore separated calibration of the bare actuator from characterization of the mechanically loaded hBN (Fig. 2). Tracking the in-plane microelectromechanical displacement during loading and unloading reveals reproducible voltage-controlled motion that largely recovers when the drive is removed (Fig. 2a). The free displacement follows $u_0(V) = \alpha V^2$ with $\alpha$=0.00128 μm/$V^2$, corresponding to approximately 1.15 μm at 30 V. This calibration defines the available electromechanical input, but not the actual hBN flake extension, which is additionally set by mechanical load sharing and interfacial transfer. The flake spans the fixed and movable sections, such that its central region bridges the actuator gap while both ends remain adhesion-coupled to the substrate (Fig. 2b,c). Optical microscopy and AFM show that the hBN long axis is oriented at $\theta \approx 17°$ relative to the actuation direction and that the thickness of the flake is approximately 43-45 nm (Fig. 2d). This tilted suspended geometry defines the mechanical transfer pathway: in-plane actuator motion is projected onto the long axis. At the same time, the adhesion boundaries constrain the transferred deformation. The PL spectra evolve systematically under microactuation, and the integrated emission intensity increases approximately quadratically with voltage and the spectral maximum blue-shift by 40.7 nm, from 844.0 to 803.3 nm (Fig. 2e,f). Its $V^2$ scaling co-evolves with the electrostatic actuator output, providing an optical response consistent with a mechanically transferred perturbation and with previously reported strain-sensitive PL from hBN defect systems[52-54].

To directly probe the lattice response, we tracked the characteristic hBN Raman mode during electrical loading processes (Fig. 2g). At zero voltage, the line-scan peak position remains nearly constant around 1366.7 $cm^{-1}$, indicating a comparatively uniform initial lattice state over the probed region (inset of Fig. 2h). Actuation produces a progressive Raman redshift that reaches approximately 9.6 $cm^{-1}$ at 30 V. Below 10 V, the peak displacement remains small relative to the uncertainty associated with small-shift Raman inversion, limiting reliable quantitative strain extraction in this regime.

Once a resolvable shift emerges, the measured peak displacement can be converted using reported Raman strain coefficients ($K_R$) into a coefficient-dependent strain window of approximately 0.07-0.14% at 10 V and 0.28-0.55% at 30 V (Fig. 2i and Supplementary Table 1)[55-57]. These values are reported as ranges rather than absolute strains due to the inversion depends on the adopted coefficient, flake thickness, local strain configuration, and interfacial transfer efficiency. Replotting the PL intensity against this Raman-inferred coordinate reveals a monotonic optical evolution with increasing lattice deformation. However, the broad coefficient-dependent intervals, particularly in the low-strain regime, limit precise discrimination of the smallest mechanically programmed states. Together with the displacement calibration and coupling geometry, the Raman and PL measurements establish hBN lattice deformation as the intermediate mechanical state through which electrical actuation is transferred to the quantum-defect host. We therefore next examine whether these mechanically programmed lattice states can be resolved reproducibly in the spin-frequency domain.

## Quantum readout of mechanical-spin modulation

To interrogate the mechanically written lattice state, we implemented an on-chip ODMR configuration in which the mechanically active hBN region is positioned adjacent to the microwave strip line (Fig. 3a-c and Supplementary Fig. 9). This arrangement preserves optical access while enabling microelectromechanical actuation and microwave spin control within the same local region. PL mapping acquired with a home-built confocal microscope confirms optically addressable $V_B^-$ emission in spatial overlap with the microwave-addressable area (Fig. 3d). The actuator was subjected to four repeated stepped loading-unloading cycles, with the voltage increased from 0 to 30 V and subsequently decreased to 0 V in 5 V increments. Across the full ODMR landscape, both resonance branches undergo reversible redshifts over each loading–unloading cycle (Supplementary Fig. 7), reproducing the programmed mechanical trajectory directly in the frequency domain (Fig. 3e). The extracted resonance center $D$ follows the same cyclic evolution and reaches an average shift of $\Delta D$= -54.80 MHz at 30 V. Similar trajectories recur over all four cycles (Fig. 3f), showing that the mechanically defined lattice state is repeatedly encoded into the spin frequencies rather than giving rise to an irreversible spectral drift.

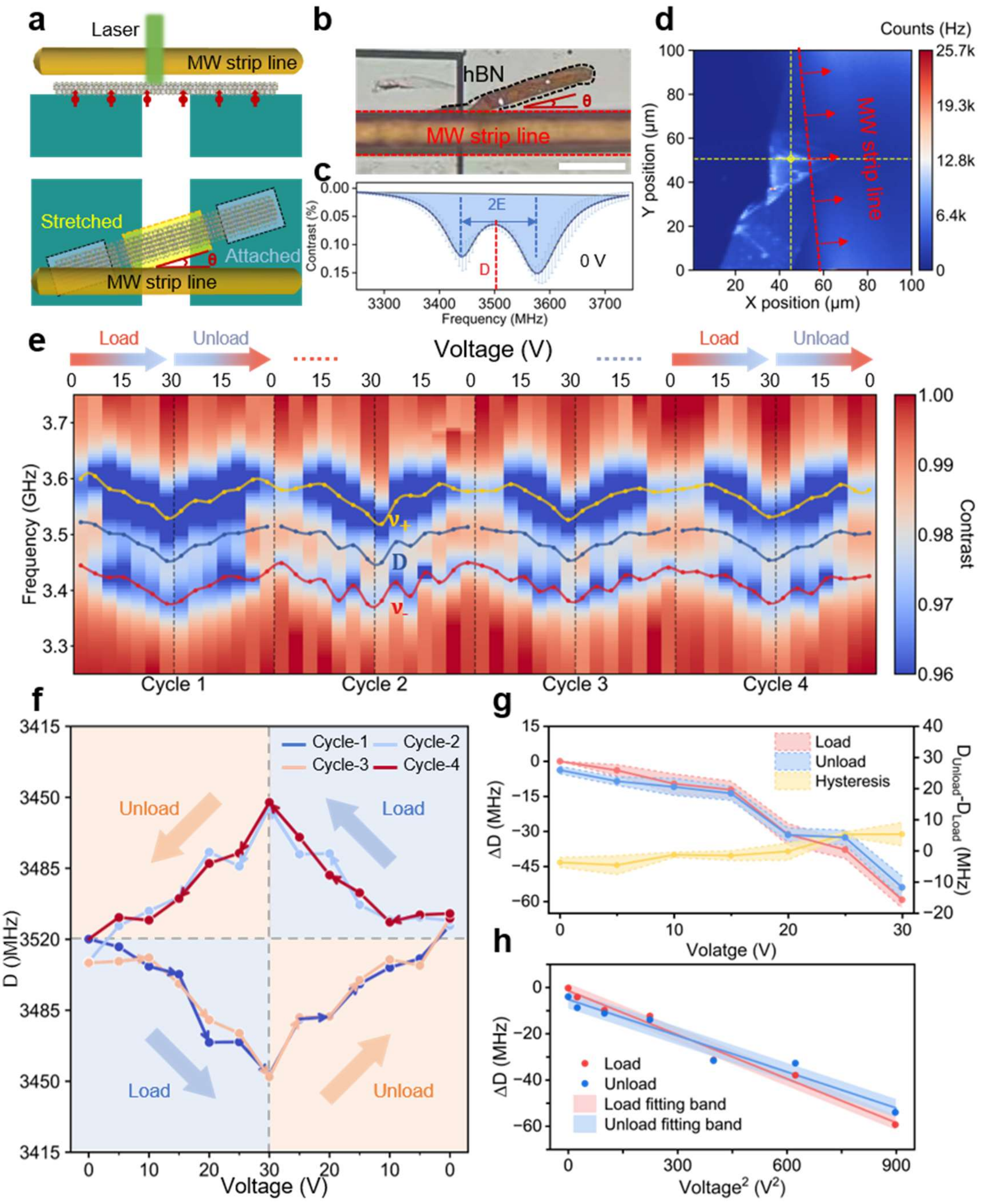

**Fig. 3 | ODMR readout of cyclic mechano-quantum modulation in OCPMQT device.**

**a-b,** Schematic and image of measurement configuration. Scale bar, 100 μm. **c,** Zero-field ODMR spectrum of the sample. **d,** 100×100 μm$^2$ confocal PL mapping of hBN flake with 1-mW laser excitation at 532 nm. The measured region and microwave transmission line are indicated. **e,** ODMR spectrum of four consecutive loading-unloading cycles, with periodic shifts in two resonant frequencies. **f,** Extracted cycle-resolved evolution of $D$, showing reproducible voltage-dependent modulation of the $V_B^-$ spin resonances by micromechanical actuation. **g,** Trajectories of the $\Delta D(V)$ $=D(V)-D(0)$, revealing path-dependent responses and partial hysteresis during cyclic actuation. **h,** Approximate linear relation between $\Delta D(V)$ and $V^2$ with fitting bands, which are consistent with the electromechanical model $F_{elec} \propto V^2$.

The $D$ response is nevertheless not perfectly single-valued with voltage. Loading and unloading follow distinct trajectories at the same applied voltage, revealing a path-dependent component in the spin-frequency output (Fig. 3f). We quantify this hysteresis as $\Delta D_{Hys}$=$D_{Unload}$-$D_{Load}$. The difference remains within ±6 MHz over the measured range and reaches 5.32 MHz at 30 V (Fig. 3g). Because the separation follows the actuation path and becomes more apparent in the high-voltage regime, it is consistent with the mechanical-transfer process mentioned above, which may involve interfacial adhesion, local sliding, or strain release[58,59]. The $D$ channel therefore retains information not only about the actuation magnitude but also the evolution of mechanical transfer across the mechano-quantum interface. Reparameterizing $\Delta D$ by $V^2$ reveals approximately linear loading and unloading relations, with slopes of $\kappa_{D-V,Load}$= $6.34\times10^{-2}$ MHz·$V^{-2}$ and $\kappa_{D-V,Unload}$= $5.21\times10^{-2}$ MHz·$V^{-2}$, respectively (Fig. 3h and Supplementary Fig. 8). To distinguish mechanically transferred strain from voltage-correlated, we measured a control hBN region outside the tensile pathway. Its $D$ remained nearly unchanged during voltage loading and unloading cycles, whereas the mechanically active region exhibited a pronounced negative shift (Supplementary Fig. 9). The preservation of the electrostatic actuation scaling in the spin-frequency output indicates that the characteristic microelectromechanical input survives the intermediate mechanical-transfer step. The difference between the two slopes retains the path-dependent component identified above. Thus, $\Delta D$ is not simply an empirical voltage-correlated spectral shift; it behaves as the frequency-domain output of a mechanically driven transduction pathway. Raman spectroscopy establishes the lattice-deformation intermediate and ODMR resolves its cyclic encoding in $D$. The remaining question is whether the magnitude of this spin response can be predicted from the chip parameters rather than fitted empirically to voltage.

## Quantitative transduction from mechanical input to spin Hamiltonian

The persistence of the $V^2$ actuation signature in $D$ suggests a continuous pathway from electrical control to the spin Hamiltonian, but scaling alone does not determine the magnitude of the frequency shift. We therefore constructed a multiscale framework that explicitly links the applied voltage, mechanically transferred hBN strain, and axial ZFS response (Fig. 4). The model follows the micromechanical input from the electrostatic actuator, through geometric projection and interfacial transfer to the strain term of the $V_{\mathrm{B}}^{-}$ spin Hamiltonian.

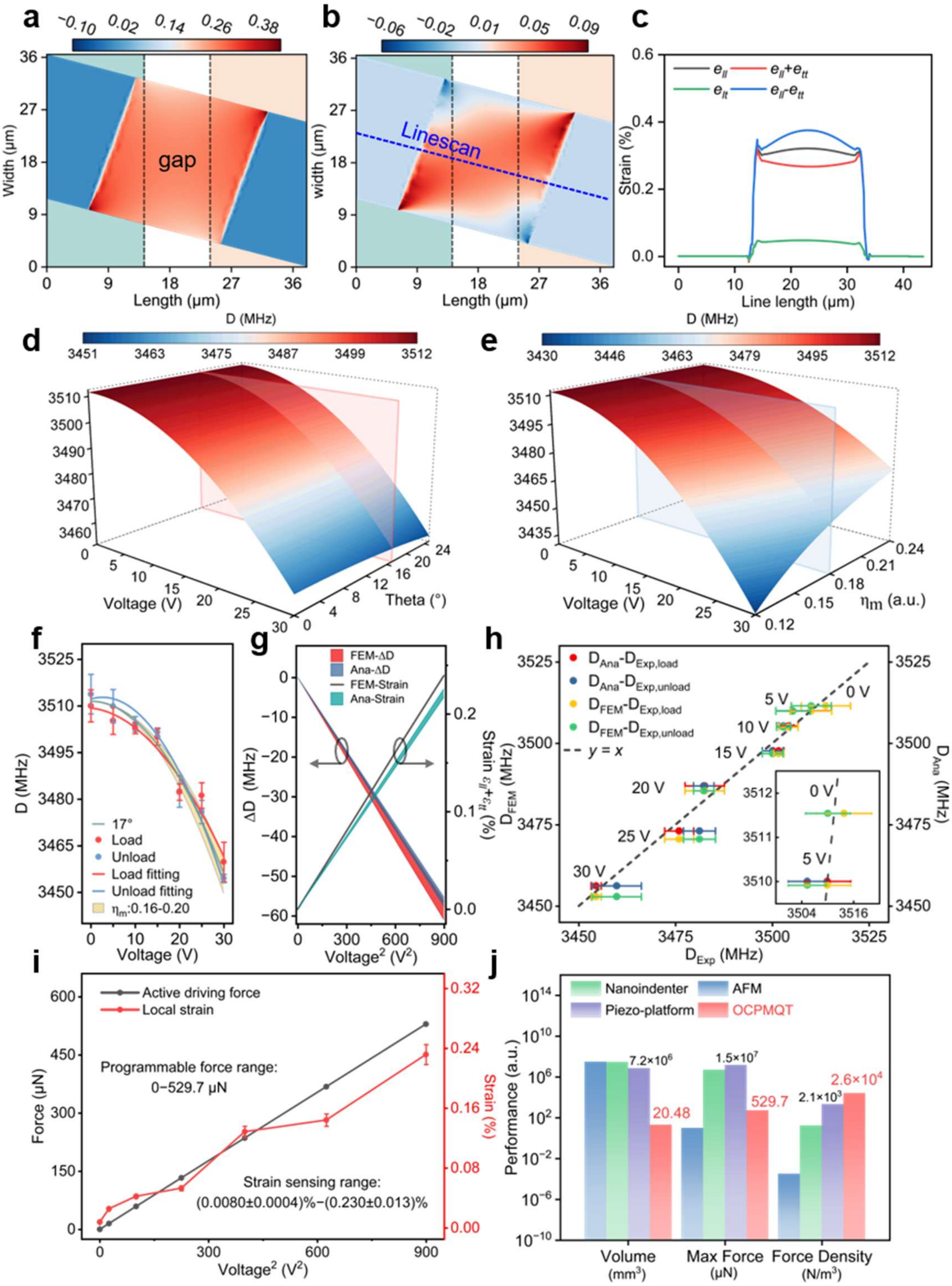

**Fig. 4 | Quantitative modeling of the voltage-dependent $D$ shift in OCPMQT.**

**a-b**, FEM simulation of the strain distribution in the tilted hBN region of OCPMQT. **a,**

Spatial distribution of the in-plane (**a**) total strain $\varepsilon_{ll}+\varepsilon_{tt}$ primarily determines the $D(V)$, and (**b**) shear strain $\varepsilon_{lt}$. **c,** multiple strain components extracted from linescan along the long axis. **d-e,** Analytical parameter surfaces showing the calculated dependence of $D$ on (**d**) voltage-tilt angle $D(V,\theta)$, and (**e**) voltage-interfacial strain-transfer efficiency $D(V,\eta_m)$. These surfaces illustrate how device geometry and strain-transfer efficiency govern the magnitude of the $D$-frequency shift. **f,** Measured and modeled analytical results of $D(V)$ with $\theta$ ≈17° and $\eta_m$: 0.16~0.20. Both loading and unloading branches exhibit a predominantly $V^2$-dependent downward shift. **g,** Comparison of the FEM-derived and analytical $\Delta D(V)$, and total strain $\varepsilon_{ll}+\varepsilon_{tt}$. **h,** Pairwise analysis of experiment, FEM simulation, and analytical modeling. The three sets of data points are distributed close to the line *y*=*x*. **i,** Force-output and strain-sensing ranges of the OCPMQT. **j,** Summary of key device metrics, including device volume, maximum output force, and volumetric force output for Nanoindentator, AFM, Piezo-platform, and OCPMQT.

Finite-element method (FEM)[48] simulations first reproduce the experimentally implemented OCPMQT geometry, including the $\theta$ ≈17° hBN orientation and in-plane micromechanical actuation (Fig. 4a,b). At 30 V, the in-plane trace strain $\varepsilon_{ll}+\varepsilon_{tt}$ is concentrated predominantly across the suspended hBN bridge, and pronounced strain gradients and shear $\varepsilon_{lt}$ develop near the mechanically coupled boundaries. These spatial features arise from the combined effects of the tilted hBN flake geometry and interfacial constraint. Linescan profiles along the hBN long axis resolve the distinct evolution of the strain components across the active region (Fig. 4c). For the $D$-channel response, the in-plane trace strain provides the first-order axial mechanical input; anisotropic and shear components primarily enter transverse spin terms and do not shift the resonance centre $D$ to first order[60]. To describe the axial strain transfer analytically, we start from the independently calibrated free actuator displacement $u_0(V) = \alpha V^2$ rather than equating this displacement with the hBN extension. Mechanical loading by the hBN flake reduces the actuator motion, imperfect interfacial coupling limits displacement transfer, and the tilted geometry projects the micromechanical motion onto the hBN long axis. Combining these effects gives the in-plane trace strain (Supplementary Note 1-3):

$$\varepsilon_{ll}^{\%}(V) + \varepsilon_{tt}^{\%}(V) = \frac{100(1+\nu_{eff})\alpha V^2 \cos\theta}{L\left[1+\left(\frac{\eta_m k_{hBN}}{Lk_x}\right)\cos^2\theta\right]}. \tag{1}$$

Here, the hBN long axis defines the local $l$ direction and the perpendicular in-plane direction defines $t$. The angle $\theta$ specifies the hBN orientation relative to micromechanical actuation; $\nu_{\mathrm{eff}}$ is the effective Poisson ratio; $k_{\mathrm{hBN}}$ and $k_x$ are the effective axial stiffnesses of the stretched hBN and actuator suspension, respectively; $L$ is the effective tensile length; and $\eta_m$ describes interfacial mechanical transfer. $\eta_m$ represents the fraction of the hBN axial rigidity engaged in device-level load sharing, rather than the intrinsic interlayer coupling of hBN. Finite transfer lengths, spatially non-uniform strain, and local interfacial slip are widely observed in supported 2D materials, with strain transmission through tens-of-nm-thick hBN occurring at the $10^{-1}$ level[58,59,61-63]. Guided by these measurements and the finite shear strength of hBN on oxygen-plasma-treated $SiO_2$, we adopt a priori range of $\eta_m$= 0.16~0.20 for the present device. Equation (1) therefore separates two physically distinct device controls: $\theta$ sets the geometric projection of micromechanical motion onto the hBN flake, whereas $\eta_m$ sets the amplitude of mechanical input transferred into the 2D lattice. The mechanically transferred strain is then mapped onto spin Hamiltonian. Neglecting hyperfine interactions for the center-frequency response, the $S$=1 ground-state spin triplet is described by[64-67]:

$$\frac{\hat{H}_0}{h} = D_0\hat{S}_z^2 + E\left(\hat{S}_x^2 - \hat{S}_y^2\right) + \frac{\hat{H}_{strain}}{h}, \tag{2}$$

where $\hat{S}_x$, $\hat{S}_y$, and $\hat{S}_z$ are dimensionless spin operators and $z$ is aligned with the crystallographic $c$ axis. The in-plane trace strain $\varepsilon_{ll} + \varepsilon_{tt}$ couples to $\hat{S}_z^2$ and shifts $D$ through $\hat{H}_{strain,z}$, whereas anisotropic and shear strains enter $\hat{H}_{strain,lt}$ through transverse quadratic spin operators. Because the present device is actuated predominantly in plane, $\varepsilon_{lz}$, $\varepsilon_{tz}$, and $\varepsilon_{zz}$ are neglected to first order. The retained strain-spin terms are[60,66]:

$$\frac{\hat{H}_{strain,z}}{h} = \left[\chi_D(\varepsilon_{ll} + \varepsilon_{tt}) + b\varepsilon_{zz}\right]\hat{S}_z^2, \tag{3}$$

$$\frac{\hat{H}_{strain,lt}}{h} = [\frac{c}{2}(\varepsilon_{ll} - \varepsilon_{tt})](\hat{S}_y^2 - \hat{S}_x^2)] + c\varepsilon_{lt}(\hat{S}_x\hat{S}_y + \hat{S}_y\hat{S}_x). \tag{4}$$

For the approximately 44-nm-thick multilayer hBN used here, we adopt an axial spin-strain coefficient $\chi_D$= -245±10 MHz/%[60].The transverse coupling coefficients are substantially smaller and are neglected, the $D(V)$ model therefore retains the axial strain contribution that controls the measured center-frequency shift. The Hamiltonian relevant to the present analysis reduces to:

$$\frac{\hat{H}(V)}{h} = D(V)\hat{S}_z^2 + E\left(\hat{S}_x^2 - \hat{S}_y^2\right). \tag{5}$$

Thus, the micromechanical-induced in-plane trace strain is encoded directly into the axial ZFS parameter. Combining the microelectromechanical input, mechanical-transfer relation, and axial strain-spin coupling gives the closed-form transduction relation (Supplementary Note 4, 5):

$$D(V) = D_0 + \chi_D \frac{100(1+\nu_{eff})\alpha V^2 \cos\theta}{L\left[1 + \left(\frac{\eta_m k_{hBN}}{L k_x}\right)\cos^2\theta\right]}. \tag{6}$$

The calculated $D(V,\theta)$ and $D(V,\eta_m)$ surfaces expose the distinct roles of device geometry and interfacial coupling (Fig. 4d,e). Increasing $\theta$ reduces the longitudinal projection of the micromechanical motion, whereas $\eta_m$ controls the magnitude of strain written into hBN. Using the measured $\theta \approx 17°$ and an interfacial-transfer window $\eta_m$: 0.16~0.20, the analytical model reproduces the measured decrease in $D$ over the full actuation range (Fig. 4f). At 30 V, the measured $\Delta D_{Exp}$ values are -59.34 MHz during loading and -50.26 MHz during unloading; both lie within the model-predicted interval of -61.88 to-50.08 MHz. The experimental mean, -54.80 MHz, differs by only 0.56 MHz (~1.0%) from the analytical prediction of -55.36 MHz at $\theta \approx 17°$. For the first time, this agreement is obtained using the independently calibrated micromechanical displacement and measured flake geometry rather than an empirical polynomial fit to the ODMR data.

The voltage scaling provides an independent test of the transduction chain. Both the FEM-derived and analytical trace strains increase approximately linearly with $V^2$, and the corresponding $\Delta D$ retains the same scaling (Fig. 4g). The quadratic electrostatic actuation law is therefore propagated through the micromechanical interface and remains measurable in the spin-frequency output. Cross-comparison of experimental, FEM-derived, and analytically predicted $D$ values places the three results close to the $y = x$ reference over the measured range (Fig. 4h). Relative to the loading experiment, the analytical and FEM-derived results yield root-mean-square errors (RMSE) of 2.74 and 3.13 MHz, respectively. The larger deviations above 20 V coincide with the regime in which path dependence becomes more pronounced, consistent with interfacial hysteresis, local sliding, or strain-release processes that are not included in the first-order model[52,58,61]. This close correspondence demonstrates that the closed-form $D(V)$ model quantitatively captures both the voltage dependence and the absolute magnitude of the spin-frequency response. By mapping the electrical control input through explicit electromechanical, geometric-projection, interfacial-transfer, and strain-spin coupling terms onto the predicted $D$ shift, the model converts the observed voltage-spin frequency correlation into a quantitative and predictive mechano-quantum transduction relation, providing the quantitative foundation of the OCPMQT.

With the transduction relation calibrated, the mechanical operating envelope of the OCPMQT can be evaluated at both the device and spin-readout levels. The programmed force and ODMR-inferred strain both increase with $V^2$, providing a continuously addressable force range from 3.6 μN to 529.7 μN and a strain-readout range from 0.0080±0.0004% to 0.230±0.013% (Fig. 4i). The correlated evolution of these quantities defines an operating window in which a on-site mechanical input can be actively applied and quantified through the $D$ channel within the same on-chip architecture.

System-level benchmarking further places this operating window in the context of conventional mechanical manipulation platforms (Fig. 4j and Supplementary Table 2). Within a device volume of $2.05\times10^{-2}$ cm$^3$, the OCPMQT delivers a maximum force of 529.7 μN, corresponding to a volumetric force output of approximately $2.6\times10^4$ N m$^{-3}$. Piezo-driven and indentation systems can provide larger absolute forces, but generally rely on substantially larger mechanical loading assemblies. The OCPMQT instead concentrates programmable sub-millinewton actuation within a chip-scale device and removes the external mechanical loading stage from the actuation pathway. Moreover, micromechanical input generation and quantum-frequency readout are co-located: the same architecture applies the perturbation, on-site transfers it to a 2D defect host, and in situ resolves the resulting $D$ shift. More broadly, programmable mechano-quantum transduction can establish a generalized architecture for force, acoustic, vibrational, and inertial sensing, interfacial micro-/nano-mechanics, and programmable spin-frequency control. This on-chip integration distinguishes the OCPMQT from a passive and discrete strain probe and establishes it as a high-performance programmable mechano-quantum transducer.

## Discussion

We have established a chip-scale framework for electrically programming the mechanical state of a solid-state quantum material and resolving its evolution through the spin Hamiltonian. Within a device volume of $2.05\times10^{-2}$ $cm^3$, the OCPMQT integrates voltage-defined micromechanical actuation, interfacial strain transfer and in situ spin-frequency readout. It delivers a maximum programmable volumetric force density of approximately $2.60\times10^4$ $N\cdot cm^{-3}$, accesses mechanically programmed states spanning ODMR-inferred strains from 0.0080±0.0004% to 0.230±0.013%, and produces a reproducible modulation of approximately -55 MHz in the ZFS parameter $D$. This combination co-localizes programmable mechanical actuation, small-strain quantum readout and spin-Hamiltonian control within a compact on-chip device architecture.

We further establish a multiscale mechano-quantum model that links micromechanical actuation, geometric projection, interfacial strain transfer, and strain-spin coupling. The resulting closed-form $D(V)$ relation reproduces the characteristic $V^2$-dependence and predicts the mean $D$ shift at 30 V, differing from the experimental value by only 0.56 MHz (~1.0%). This framework transforms the observed voltage-spin frequency correlation into a predictive transduction relation, enabling quantitative attribution of device mechanics extension to other mechanically responsive quantum materials. The OCPMQT advances solid-state quantum sensing from spatially separated and passive strain readout towards programmable on-chip mapping of mechanical inputs onto the spin Hamiltonian. Its on-chip mechanical-programming and quantum-readout architecture provides a device-level foundation for compact, multimodal and integrable quantum systems, supporting the transition from laboratory-scale measurements to deployable quantum technologies.

## Author contributions

T.Y. proposed the concept of the on-chip programmable mechano-quantum transducer. X.Z. and T.Y. were in charge of the OCPMQT system design and simulation, chip fabrication and test, data acquisition and results analysis. X.Z., W.L., N.G., Z.M., and T.Y. were in charge of chip optical test, data acquisition, and results analysis. L.X contributed to optical test, data acquisition, and results analysis. D.M., Z.G., Y.W., J.X., X.L., and Z.M. contributed to the chip fabrication and test. All authors contributed to the experimental analysis and interpretation of results. X.Z., W.L., and T.Y. wrote the paper with input from all authors. C.L., G.G., J.T., and T.Y. supervised the whole project.

## Competing interests

The authors declare no competing interests.